\documentclass[english,12pt]{article}
\usepackage[cp1251]{inputenc}
\usepackage{babel}
\usepackage{amsfonts, amsmath}
\newcommand{\be}{\begin{equation}} \newcommand{\ee}{\end{equation}}

\begin{document}
\title{Generalized Uncertainty Relations,Fundamental Length and Density
Matrix } \thispagestyle{empty}

\author{A.E.Shalyt-Margolin \hspace{1.5mm}$^1$\thanks
{Phone (+375) 172 883438; e-mail: a.shalyt@mail.ru; alexm@hep.by
},A.Ya.Tregubovich$^2$
\thanks{Phone (+375) 172 841559
; e-mail: a.tregub@open.by} }
\date{}
\maketitle
 \vspace{-25pt}
{\footnotesize\noindent $^1$National Centre of High Energy and
Particle Physics, Bogdanovich Str.153, Minsk 220040, Belarus\\
$^2$Institute of Physics National Academy of Sciences
                          Skoryna av.68, Minsk 220072, Belarus}\\

\vspace{0.25cm}
 {\ttfamily{\footnotesize

 }}

\rm\normalsize \vspace{0.25cm}
\begin{abstract}
It was shown that if in Quantum Theory a fundamental length exists
and a well-known measurement procedure is used, then the density
matrix at the Planck scale cannot be defined in the usual way,
because in this case density matrix trace is strongly less than
one.
 Density matrix must be changed by a progenitrix or as we call it
throughout this paper, density pro-matrix. This pro-matrix is a
deformed density matrix, which at low energy limit turns to usual
one.
 Below the explicit form of the deformation is described.
Implications of obtained results are summarized as well as their
application to the interpretation of Information Paradox
 on the Black Holes.
 \end{abstract}
 \newpage
\section{Introduction}
In this paper
 we will show that if in Quantum Theory there is a
fundamental length and a known measurement procedure is used, then
the density matrix at the Planck scale cannot be defined in the
usual way, because in this case density matrix trace is strongly
less than one.Since from Generalized Uncertainty Relations (GUR)
follow that there is a fundamental length, then one of the
implications of existence of GUR is the fact that in this case
density matrix  cannot be defined in the usual way. It was shown,
that commonly accepted definition of density matrix cannot be used
at Planck scale and it is necessary to use density pro-matrix,
which appears when Quantum Mechanics with Fundamental Length
(QMFL) is considered as a deformation usual Quantum Mechanics
(QM). This deformation is described explicitly. It was shown
also,that inflationary model contains two different (unitary
non-equivalent) Quantum Mechanics: the first one describes nature
at the Planck scale and it is QMFL. The second one is obtained as
a limit transition from Planck scale to low energy one and it is
based on the Heisenberg Uncertainty Relations (UR) \cite{r1}.It is
QM. The interpretation of obtained results as well as their
implications are discussed below, in particular for explaining the
information paradox in primordial black holes .

\section{General Uncertainty Relations,
 Fundamental Length and Density Matrix}

Using different approaches (String Theory \cite{r2}, Gravitation
\cite{r3}, Quantum Theory of black holes \cite{r4}  etc
\cite{r5}). the authors of numerous papers issued over the last
14-15 years have pointed out that Heisenberg's Uncertainty
Relations should be modified. Specifically, a high energy
correction has to appear
\begin{equation}\label{U2}
\triangle x\geq\frac{\hbar}{\triangle p}+\alpha
L_{p}^2\frac{\triangle p}{\hbar}.
\end{equation}
\noindent Here $L_{p}$ is the Planck's length:
$L_{p}=\sqrt\frac{G\hbar}{c^3}\simeq1,6\;10^{-35}\;m$ and
 $\alpha > 0$ is a constant. In \cite{r3} it was shown
that this constant may be chosen equal to 1. However, here we will
use $\alpha$ as an arbitrary constant without giving it any
definite value. The  inequality (\ref{U2}) is quadratic in
$\triangle p$:
\begin{equation}\label{U3}
\alpha L_{p}^2({\triangle p})^2-\hbar \triangle x \triangle p+
\hbar^2 \leq0,
\end{equation}
from whence the fundamental length is
\begin{equation}\label{U4}
\triangle x_{min}=2\sqrt\alpha L_{p}.
\end{equation}
Since in what follows we proceed only from the existence of
fundamental length, it should be noted that this fact was
established apart from GUR as well. For instance, from an ideal
experiment associated with Gravitational Field and Quantum
Mechanics a lower bound on minimal length was obtained in
\cite{r6}, \cite{r7} and  improved in \cite{r8} without using GUR
to an estimate of the form $\sim L_{p}$. \noindent Let us to
consider equation (\ref{U4}) in some detail. Squaring both its
sides, we obtain
\begin{equation}\label{U5}
(\overline{\Delta\widehat{X}^{2}})\geq 4\alpha L_{p}^{2 },
\end{equation}
Or in terms of density matrix
\begin{equation}\label{U6}
Sp[(\rho \widehat{X}^2)-Sp^2(\rho \widehat{X}) ]\geq 4\alpha
L_{p}^{2 }=l^{2}_{min}>0,
\end{equation}
where $\widehat{X}$ is the coordinate operator. Expression
(\ref{U6}) gives the measuring rule used in QM. However, in the
case considered here, in comparison with QM, the right part of
(\ref{U6}) cannot be done arbitrarily near to zero since it is
limited by $l^{2}_{min}>0$, where due to GUR $l_{min} \sim L_{p}$.

 Apparently, this may be due to
the fact that QMFL with GUR (\ref{U2}) is unitary non-equivalent
to  QM with UR. Actually, in QM the left-hand side of (\ref{U6})
can be chosen arbitrary closed to zero, whereas in QMFL this is
impossible. But if two theories are unitary equivalent then, the
form of their spurs should be retained. Besides, a more important
aspect is contributing to unitary non-equivalence of these two
theories: QMFL contains three fundamental constants (independent
parameters) $G$, $c$ and $\hbar$, whereas QM contains only one:
$\hbar$. Within an inflationary model (see \cite{r10}), QM is the
low-energy limit of QMFL (QMFL turns to QM) for the expansion of
the Universe. In this case, the second term in the right-hand side
of (\ref{U2}) vanishes and GUR turn to UR. A natural way for
studying QMFL is to consider this theory as a deformation of QM,
turning to QM at the low energy limit (during the expansion of the
Universe after the Big Bang). We will consider precisely this
option. However differing from authors of papers \cite{r4},
\cite{r5} and others, we do not deform commutators, but density
matrix, leaving at the same time the fundamental
quantum-mechanical measuring rule (\ref{U6}) without changes. Here
the following question may be formulated: how should be deformed
density matrix conserving quantum-mechanical measuring rules in
order to obtain self-consistent measuring procedure in QMFL? For
answering to the question we will use the R-procedure. For
starting let us to consider R-procedure both at the Planck's
energy scale and at the low-energy one. At the Planck's scale $a
\approx il_{min}$ or $a \sim iL_{p}$, where $i$ is a small
quantity. Further $a$ tends to infinity and we obtain for density
matrix $$Sp[\rho a^{2}]-Sp[\rho a]Sp[\rho a] \simeq
l^{2}_{min}\;\; or\;\; Sp[\rho]-Sp^{2}[\rho] \simeq
l^{2}_{min}/a^{2}.$$

 Therefore:

 \begin{enumerate}
 \item When $a < \infty$, $Sp[\rho] =
Sp[\rho(a)]$ and
 $Sp[\rho]-Sp^{2}[\rho]>0$. Then, \newline $Sp[\rho]<1$
 that corresponds to the QMFL case.
\item When $a = \infty$, $Sp[\rho]$ does not depend on $a$ and
$Sp[\rho]-Sp^{2}[\rho]\rightarrow 0$. Then, $Sp[\rho]=1$ that
corresponds to the QM case.
\end{enumerate}
How should be points 1 and 2 interpreted? How does analysis
above-given agree to the main result from \cite{r21} \footnote
{"... there cannot be any physical state which is a position
eigenstate since a eigenstate would of course have zero
uncertainty in position"}? It is in full agreement. Indeed, when
state-vector reduction (R-procedure) takes place in QM then,
always an eigenstate (value) is chosen exactly. In other words,
the probability is equal to 1. As it was pointed out in the
above-mentioned point 1 the situation changes when we consider
QMFL: it is impossible to measure coordinates exactly since it
never will be absolutely reliable. We obtain in all cases a
probability less than 1 ($Sp[\rho]=p<1$). In other words, any
R-procedure in QMFL leads to an eigenvalue, but only with a
probability less than 1. This probability is as near to 1 as far
the difference between measuring scale $a$ and $l_{min}$ is
growing, or in other words, when the second term in (\ref{U2})
becomes insignificant and we turn to QM. Here there is not a
contradiction with \cite{r21}. In QMFL there are not exact
coordinate eigenstates (values) as well as there are not pure
states. In this paper we do not consider operator properties in
QMFL as it was done in \cite{r21} but density-matrix properties.

 The  properties of density matrix in
QMFL and QM have to be different. The only reasoning in this case
may be as follows: QMFL must differ from QM, but in such a way
that in the low-energy limit a density matrix in QMFL must
coincide with the density matrix in QM. That is to say, QMFL is a
deformation of QM and the parameter of deformation depends on the
measuring scale. This means that in QMFL $\rho=\rho(x)$, where $x$
is the scale, and for $x\rightarrow\infty$  $\rho(x) \rightarrow
\widehat{\rho}$, where $\widehat{\rho}$ is the density matrix in
QM.

Since on the Planck's scale $Sp[\rho]<1$, then for such scales
$\rho=\rho(x)$, where $x$ is the scale, is not a density matrix as
it is generally defined in QM. On Planck's scale we name $\rho(x)$
 "density pro-matrix". As follows from the above, the density
matrix $\widehat{\rho}$ appears in the limit
\begin{equation}\label{U12}
\lim\limits_{x\rightarrow\infty}\rho(x)\rightarrow\widehat{\rho},
\end{equation}
when GUR (\ref{U2}) turn to UR  and QMFL turns to QM.

Thus, on Planck's scale the density matrix is inadequate to obtain
all information about the mean values of operators. A "deformed"
density matrix (or pro-matrix) $\rho(x)$ with $Sp[\rho]<1$ has to
be introduced because a missing part of information $1-Sp[\rho]$
is encoded in the quantity $l^{2}_{min}/a^{2}$, whose specific
weight decreases as the scale $a$ expressed in  units of $l_{min}$
is going up.

\section{QMFL as a deformation of QM}
Here we are going to describe QMFL as a deformation of QM using
the density pro-matrix formalism. In this context density
pro-matrix has to be understood as a deformed density matrix in
QMFL. As fundamental deformation parameter we will use
$\beta=l_{min}^{2}/x^{2 }$, where $x$ is the scale.

\noindent {\bf Definition 1.}

\noindent Any system in QMFL is described by a density pro-matrix
$\rho(\beta)=\sum_{i}\omega_{i}(\beta)|i><i|$, where
\begin{enumerate}
\item $0<\beta\leq1/4$;
\item The vectors $|i>$ form a full orthonormal system;
\item $\omega_{i}(\beta)\geq 0$ and for all $i$ there is a
finite limit $\lim\limits_{\beta\rightarrow
0}\omega_{i}(\beta)=\omega_{i}$;
\item
$Sp[\rho(\beta)]=\sum_{i}\omega_{i}(\beta)<1$,
$\sum_{i}\omega_{i}=1$;
\item For any operator $B$ and any $\beta$ there is a
 mean operator $B$, which depends on  $\beta$:\\
$$<B>_{\beta}=\sum_{i}\omega_{i}(\beta)<i|B|i>.$$
\end{enumerate}
At last, in order to match our definition with the result of
section 2 the next condition has to be fulfilled:
\begin{equation}\label{U13}
Sp[\rho(\beta)]-Sp^{2}[\rho(\beta)]\approx\beta,
\end{equation}
from which we can find the meaning of the quantity
$Sp[\rho(\beta)]$, which satisfies the condition of definition:
\begin{equation}\label{U14}
Sp[\rho(\beta)]\approx\frac{1}{2}+\sqrt{\frac{1}{4}-\beta}.
\end{equation}

From point 5. it follows, that $<1>_{\beta}=Sp[\rho(\beta)]$.
Therefore for any scalar quantity $f$ we have $<f>_{\beta}=f
Sp[\rho(\beta)]$. In particular, the mean value
$<[x_{\mu},p_{\nu}]>_{\beta}$ is equal to
\begin{equation}\label{U15}
<[x_{\mu},p_{\nu}]>_{\beta}= i\hbar\delta_{\mu,\nu}
Sp[\rho(\beta)]
\end{equation}
We will call density matrix the limit
$\lim\limits_{\beta\rightarrow 0}\rho(\beta)=\rho$. It is evident,
that in the limit $\beta\rightarrow 0$ we turn to QM. Here we
would like to verify, that two cases described above correspond to
the meanings of $\beta$. In the first case when $\beta$ is near to
1/4. In the second one when it is near to zero.
\\
From the definitions given above it follows that
$<(j><j)>_{\beta}=\omega_{j}(\beta)$. From which the condition of
completeness on $\beta$ is
\\$<(\sum_{i}|i><i|)>_{\beta}=<1>_{\beta}=Sp[\rho(\beta)]$. The
norm of any vector $|\psi>$, assigned to  $\beta$ can be defined
as
\\$<\psi|\psi>_{\beta}=<\psi|(\sum_{i}|i><i|)_{\beta}|\psi>
=<\psi|(1)_{\beta}|\psi>=<\psi|\psi> Sp[\rho(\beta)]$, where
$<\psi|\psi>$ is the norm in QM, or in other words when
$\beta\rightarrow 0$. By analogy, for probabilistic interpretation
the same situation takes place in the described theory, but only
changing $\rho$ by $\rho(\beta)$.
\\

\renewcommand{\theenumi}{\Roman{enumi}}
\renewcommand{\labelenumi}{\theenumi.}
\renewcommand{\labelenumii}{\theenumii.}

Some remarks:

\begin{enumerate}
\item The considered above limit covers at the same time
Quantum and Classical Mechanics. Indeed, since
$\beta=l_{min}^{2}/x^{2 }=G \hbar/c^3 x^{2 }$, so we obtain:
\begin{enumerate}
\item $(\hbar \neq 0,x\rightarrow
\infty)\Rightarrow(\beta\rightarrow
0)$ for QM;
\item $(\hbar\rightarrow 0,x\rightarrow
\infty)\Rightarrow(\beta\rightarrow
0)$ for Classical Mechanics;
\end{enumerate}
\item The parameter of deformation $\beta$
should take the meaning $0<\beta\leq1$. However, as we can see
from (\ref{U14}), and as it was indicated in the section 2,
$Sp[\rho(\beta)]$ is well defined only for $0<\beta\leq1/4$.That
is if $x=il_{min}$ and $i\geq 2$ then, there is not any problem.
 At the very point with fundamental
length $x=l_{min}\sim L_{p}$ there is a singularity, which is
connected with the appearance of the complex value  of
$Sp[\rho(\beta)]$, or in other words it is connected with the
impossibility of obtain a diagonalized density pro-matrix at this
point over the field of real numbers. For this reason definition 1
at the initial point do not has any sense.
\item We have to consider the question about solutions
(\ref{U13}). For instance, one of the solutions (\ref{U13}), at
least at first order on $\beta$ is
$\rho^{*}(\beta)=\sum_{i}\alpha_{i} exp(-\beta)|i><i|$, where all
$\alpha_{i}>0$ do not depend on $\beta$  and their sum is equal to
1, that is $Sp[\rho^{*}(\beta)]=exp(-\beta)$. Indeed, we can easy
verify that
\begin{equation}\label{U15}
Sp[\rho^{*}(\beta)]-Sp^{2}[\rho^{*}(\beta)]=\beta+O(\beta^{2}).
\end{equation}
\item It is clear, that in the proposed description of
states, which have a probability equal to 1, or in others words
pure states can appear only in the limit $\beta\rightarrow 0$, or
when all states $\omega_{i}(\beta)$ except one of them are equal
to zero, or when they tend to zero at this limit.
\item We suppose, that all definitions concerning
density matrix can be transferred to the described above
deformation of Quantum Mechanics (QMFL) changing the density
matrix $\rho$ by the density pro-matrix $\rho(\beta)$ and turning
then to the low energy limit $\beta\rightarrow 0$. In particular,
for statistical entropy we have
\begin{equation}\label{U16}
S_{\beta}=-Sp[\rho(\beta)\ln(\rho(\beta))].
\end{equation}
The quantity $S_{\beta}$, evidently never is equal to zero, since
$\ln(\rho(\beta))\neq 0$ and, therefore $S_{\beta}$ may be equal
to zero only at the limit $\beta\rightarrow 0$.
\end{enumerate}

\renewcommand{\theenumi}{\arabic{enumi}}

\section{Some Implications}
\begin{enumerate}
\item If we carry out a measurement in a defined scale, we cannot
consider a density pro-matrix  with a precision, which exceed some
limit of order $\sim10^{-66+2n}$ , where $10^{-n}$ is the scale in
which the measurement is carried out. In most of the known cases
this precision is quite enough for considering density pro-matrix
the density matrix. However, at the Planck scale, where Quantum
Gravity effects cannot be neglected and energy is of the Planck
order the difference between $\rho(\beta)$ and $\rho$ have to be
considered.
\item At the Planck scale the notion of Wave Function of the
Universe, introduced by J.A. Wheeler and B. deWitt \cite{r9} does
not work and in this case quantum gravitation effects can be
described only with the help of density pro-matrix $\rho(\beta)$.
\item Since density pro-matrix $\rho(\beta)$ depends on the scale in which
the measurement is carried out, so the evolution of the Universe
within inflation model paradigm \cite{r10} is not an unitary
process, because, otherwise the probability
$p_{i}=\omega_{i}(\beta)$   would be conserved.
\end{enumerate}
\section{On the problem of information paradox in Black Holes}
The results obtained above give us the opportunity for considering
again the problem of loss information on Black Holes
\cite{r11,r12,r13}, at least for the case of primordial Black
Holes. Indeed, because when we consider these Black Holes the
Planck's scale is important, then as it was shown above the
entropy of matter  observed by a Black Hole at this scale is not
equal to zero, as it was confirmed by R. Myers \cite{r14}.
According to his results a pure state cannot form a Black Hole. In
this case it is necessary to reformulate the problem itself, since
in all published papers on information paradox up to now the equal
to zero entropy at the initial state is equal to  non zero one at
the final state. It is necessary to note, that last time some
papers have been issued, where QM with GUR is considered at the
very beginning. As a consequence of this approach an stable
remnants due to the process of Black Hole evaporation appears.

On the other hand from results obtained above, qualitatively we
can answer to the question about information loss on the black
holes, which are formed when a star collapses. Indeed, near to the
horizon of events an approximately pure state has an entropy
practically equal to zero: $S^{in}=-Sp[\rho \ln(\rho)]$, which
corresponds to the value $\beta \to O$. When it is approaching to
a singularity $\beta>0$ (in other words to the Plank scale) and it
has yet a non equal to zero entropy: $S_{\beta}=-Sp[\rho
(\beta)\ln(\rho(\beta))]$. Therefore in a black hole entropy
increases as well as information is lost.

\section{Conclusion}
There is a question. Is it rightful to use the commonly defined
measurement procedure in Quantum Gravity? So far in many papers on
Quantum gravity (see for instance \cite{r16}) any other procedure
has not be used or proposed. But as it was shown above in the case
when Quantum Gravity effects are important there are not pure
states. And the other hand as it was noted in \cite{r17} all known
approaches to justify Quantum Gravity one way or another lead to
the notion of fundamental length. Besides that GUR (\ref{U2}),
which as well lead to that notion are well described within the
inflation model \cite{r18}. Therefore, apparently is not possible
to understand physics at the Planck's scale without these notions.
Besides that, it is necessary to consider one more aspect of this
problem. As it was noted in \cite{r19}, when a new physical theory
is created, it implies the introduction of a new parameter and the
deformation of precedent theory by this parameter. All these
deformation parameters are in their essence fundamental constants:
$G$, $c$ and $\hbar$ (more exactly in \cite{r19} instead of $c$,
$1/c$ is used). From the results presented above it follows, that
the question formulated in \cite{r19} can be specified: is it this
theory, the theory with fundamental length, which contains by
definition all these three parameters:
$L_{p}=\surd\frac{G\hbar}{c^3}$ ?
\\This paper is the extended and revised version of \cite{r20}.



\end{document}